\documentclass[11pt,a4paper]{article}
\usepackage{amsthm}
\usepackage{amsmath}
\usepackage{amscd}
\usepackage[latin2]{inputenc}
\usepackage{t1enc}
\usepackage[mathscr]{eucal}
\usepackage{indentfirst}
\usepackage{graphicx}
\usepackage{graphics}
\usepackage{pict2e}
\usepackage{epic}
\numberwithin{equation}{section}
\usepackage[margin=3cm]{geometry}
\usepackage{epstopdf} 
\usepackage{bbold}
\usepackage{amsfonts}
\usepackage{amssymb,tikz}
\usepackage{enumerate}
\usepackage{hyperref}

\theoremstyle{plain}
\newtheorem{thm}{Theorem}

\theoremstyle{definition}
\newtheorem{defn}[thm]{Definition}

\theoremstyle{remark}

\def\FF{\mathbb{F}}
\def\RR{\mathbb{R}}
\def\ZZ{\mathbb{Z}}
\def\FFF{\mathcal{F}}
\def\PP{\mathcal{P}}

\def\SS{\mathcal{S}}
\def\TT{\mathcal{T}}

\def\ntt{\mathrm{NTT}}
\def\HB{\mathrm{HB}}
\def\LB{\mathrm{LB}}

\usetikzlibrary{positioning}

\usepackage{authblk}

\begin{document}

\title{A survey on NIST PQ signatures}


\author[1]{Nicola Di Chiano}
\author[2]{Riccardo Longo}
\author[2]{Alessio Meneghetti}
\author[2]{Giordano Santilli}
\affil[1]{Department of Mathematics, University of Bari Aldo Moro}
\affil[2]{Department of Mathematics, University of Trento}

\maketitle

Shor's shockingly fast quantum algorithm for solving the period-finding problem is a threat for the most common public-key primitives, as it can be efficiently applied to solve both the Integer Factorisation Problem and the Discrete Logarithm Problem. In other words, as soon as a large-enough quantum computer is born, many once-secure protocols have to be replaced by still-secure alternatives. Instead of relying, for example, on the RSA protocol, the Diffie-Hellman key-exchange or the (Elliptic Curve) Digital Signature Algorithm, many researchers moved their attention to the design and analysis of primitives which are yet to be broken by quantum algorithms.
\\
The urgency of the threat imposed by quantum computers led the U.S. National Institute of Standards and Technology (NIST) to open calls for both Post-Quantum Public-Keys Exchange Algorithms and Post-Quantum Digital Signature Algorithms \cite{NIST-stand}. This new NIST standardisation process started in 2016, has involved hundreds of researchers, 
has seen 37 early submissions for a total of 82 proposals, and has recently reached its third round of analyses.
\\
In this brief survey we focus on the round 3 finalists and alternate candidates for Digital Signatures, announced on July 22, 2020:
\begin{center}
\begin{tabular}{p{.45\linewidth}|p{.45\linewidth}}
\hline
Finalists&Alternate Candidates\\
\hline
CRYSTALS-DILITHIUM \cite{Crystals} & SPHINCS$^+$ \cite{Sphincs+}\\
FALCON \cite{FALCON} & GeMSS \cite{GeMSS}\\
Rainbow \cite{Rainbow}&Picnic \cite{Picnic}\\
\hline
\end{tabular}
\end{center}
These schemes are designed to address distinct security levels, known as Security Level I, III and V. These levels correspond to, respectively, 128, 192 and 256 bits of security against collisions. Among the six schemes above, only Falcon cannot be instantiated to all three security levels.
In order to present these primitives we start, in Section \ref{sec: preliminaries}, with an introduction to their underlying mathematical objects and the related problems, i.e. lattices, polynomial ideals, one-way functions and zero-knowledge proofs. Then, Section \ref{sec: schemes} describes the six digital signatures and lists their algorithms for key-generation, signing and verification. Finally, in Section \ref{sec: conclusions} we conclude with a comparison between the different schemes. 
\section{Preliminaries}\label{sec: preliminaries}
\subsection{Digital Signatures}
A digital signature is a public-key protocol that acts as the digital counterpart of a traditional signature. Formally, the properties that a digital signature must achieve are the following \cite{NIST1}:
\begin{enumerate}[1)]
    \item \emph{Authentication}: the receiver of the document must be sure of the identity of the sender.  
    \item \emph{Integrity}: the signed document should not be altered when transmitted.
    \item \emph{Non-repudiation}: the signer of the document cannot deny having signed the document.
    \item \emph{Non-reusability}: the signature must be used only once.
    \item \emph{Unforgeability}: only the signer of the message should be able to give a valid signature.
\end{enumerate}
A signature scheme is usually composed by three algorithms: 
\begin{itemize}
    \item[$\bullet$] \emph{Keys Generation Algorithm}: using the global parameters defined at the beginning of the scheme, this algorithm generates a private key and the corresponding public key.
    \item[$\bullet$] \emph{Signing Algorithm}: using the private key and the message needed to be signed, this algorithm outputs a signature. 
    \item[$\bullet$] \emph{Verification Algorithm}: using the public key and the message, the receiver is able to decide whether the signature obtained is valid or not. 
\end{itemize}

A detailed description of digital signature schemes can be found in \cite{Stinson, Menezes}.

\subsection{Lattice Theory}
\begin{defn}
Let $m$ be a positive integer. A discrete additive subgroup $L$ of $\RR^m$ is called a (Euclidean) lattice. An equivalent definition may be given in terms of linear algebra: given a finite set of linearly independent vectors $B=\{v_1, \ldots, v_n\}$ of $\RR^m$, a lattice $L(B)$ is the set of the linear combinations with integer coefficients of the set B. The set $B$ is called a basis of the lattice $L$ and $n$ is the dimension of the lattice. It is possible to write a $n\times m$ matrix $A$ associated to a lattice, in which the rows of the matrix are the coordinates of the vectors of the basis.
\end{defn}
In order to expose the most famous lattice problems, on which lattice cryptography is based, we need to introduce the  minimal distance $\lambda_1(L)$, that is $\lambda_1(L)=\min_{v \in L \setminus \{0\}} ||v||$. Analogously it is possible to define the successive minimum $\lambda_i(L)$ for $2 \leq i \leq n$ as $$\lambda_i(L)=\min_r \{v_1, \ldots, v_i \text{ independent in } L  \::\: ||v_j|| \leq r \, \text{for } 1\leq j \leq i  \}.$$ 
Moreover we need to consider other algebraic structures: given a polynomial $\phi(x) \in \ZZ[x]$, usually $\phi(x)=x^n-1$ or $\phi(x)=x^n+1$, and a prime $q > 2$, we define $R=\ZZ[x]/\left(\phi(x)\right)$ and $R_q=R/qR=\ZZ_q[x]/\left(\phi(x)\right)$. 

Some famous examples of hard lattice problems are the following:
\begin{itemize}
    \item[$\bullet$] 
    (SVP) Given a basis $B$ of a lattice $L$, find a vector $v \in L$ such that $||v||=\lambda_1(L)$.\\
    (Approx-SVP$_\gamma$) Given a basis $B$ of a lattice $L$ find a vector $v \in L$ such that $||v||\leq \gamma(n) \lambda_1(L)$, where the constant $\gamma(n)$ depends on the dimension of the lattice $n$.\\
    (GapSVP$_\gamma$) Given a basis $B$ of a lattice $L$ and a constant $d$, decide if  $\lambda_1(L) \leq d$ or $\lambda_1(L) > \gamma d$.
    \item[$\bullet$]
    (SIVP$_\gamma$) Given a basis $B$ of a lattice $L$ find linearly independent vectors $v_1, \ldots, v_n \in L$ such that $||v_i|| \leq \gamma\lambda_n(L)$ for $1 \leq i \leq n$.
    \item[$\bullet$] 
    (SIS$_\beta$, \cite{Ajtai, MR1}) Given a matrix $A \in M_{n \times m}(\ZZ_q)$, find a vector $z \in \ZZ_q^m$ such that $Az \equiv 0 \bmod{q}$ and $||z|| \leq \beta$.\\
    (RSIS$_\beta$, \cite{PR, LM}) Given a vector $a \in R_q^m$, find a vector $z \in R_q^m$ such that $\langle z, b \rangle=0$ and $||z|| \leq \beta$.\\
    (MSIS$_\beta$, \cite{LS})  Given a matrix $A \in M_{n \times m}(R_q)$, find a vector $z \in R_q^m$ such that $Az \equiv 0 \bmod{q}$ and $||z|| \leq \beta$.
    \item[$\bullet$] 
    For a vector $s \in \ZZ_q^n$ and a discrete Gaussian distribution $\chi$, with width $\alpha q$ for some $\alpha < 1$, the LWE distribution $A_{s,\chi}$ is sampled by choosing $a \in \ZZ_q^n$ uniformly at random, $e$ drawn with $\chi$ and outputting the pair $\left(a,b=\langle a, s \rangle +e \bmod{q} \right)$. \\
   (LWE, \cite{Regev1}) Given $m$ pairs $\left(a_1,b_1 \right), \ldots, \left(a_m,b_m \right)$ drawn from $A_{s,\chi}$ for a random $s \in \ZZ_q^n$, find $s$.  \\
  For $s \in R_q$ the RLWE distribution $A_{s,\chi}^R$ is sampled by choosing $a \in R_q$ uniformly at random, $e$ drawn with $\chi$ and outputting the pair $\left(a,b=s \cdot a +e \bmod{q} \right)$. \\
   (RLWE, \cite{LPR}) Given $m$ pairs $\left(a_1,b_1 \right), \ldots, \left(a_m,b_m \right)$ drawn from $A_{s,\chi}^R$ for a random $s \in R_q$, find $s$.  \\
      For a vector $s \in R_q^n$, the MLWE distribution $A_{s,\chi}^M$ is sampled by choosing $a \in R_q^n$ uniformly at random, $e$ drawn with $\chi$ and outputting the pair $\left(a,b=\langle a, s \rangle +e \bmod{q} \right)$. \\
   (MLWE, \cite{LS, BGV})  Given $m$ pairs $\left(a_1,b_1 \right), \ldots, \left(a_m,b_m \right)$ drawn from $A_{s,\chi}^M$ for a random $s \in R_q^n$, find $s$.  
   \end{itemize}

Although it seems that the problems of the SIS and LWE families are not related to those on lattices, it can be proved \cite{Regev3, LPR, Peikert, MR2, LS} that, by using several reductions, solving these problems is as least as difficult as solving instances of GapSVP and SIVP.

See \cite{Silverman, Regev2, Peikert2} for further details on lattices and lattice cryptography.

\subsection{Multivariate polynomial systems theory}
\begin{defn}Let $m,n,q$ be three positive integers with $m\leq n$ and $\FF_q$ a finite field of cardinality $q$. Let $p_1,\ldots,p_m\in \FF_q[x_1,\ldots,x_n]$ be $m$ quadratic polynomials in $n$ variables. The MQ (multivariate quadratic) problem consists of finding a solution $\bar{x}\in {\FF_q}^n$  of the system:
\begin{equation}\label{system} p_1(x_1,\ldots,x_n)=p_2(x_1,\ldots,x_n)=\ldots=p_m(x_1,\ldots,x_n)=0. \end{equation}
\end{defn}

 When the system \eqref{system} is random, i.e. for all $i=1,\ldots,m$  the  coefficients of $p_i$ are choosen uniformly at random, MQ  has been proven to be an NP-hard problem \cite{PG}. Since a quantum algorithm for solving MQ problem does not exist, the multivariate protocols are very used in Post-Quantum cryptography.

The field $\FF_q$ is not algebraically closed, therefore a solution of \eqref{system} definitely belongs to ${\FF_q}^n$ if the field equations are added to the system. So, given the polynomial ideal $I=\langle p_1,\ldots,p_m \rangle$,  if the solution of \eqref{system} is unique, solving the MQ problem is equivalent to find the only point of the variety $V(I)\cap{\FF_q}^n=V(I+\langle x_1^q-x_1,\ldots,x_n^q-x_n \rangle )$ \cite{KR}.

In terms of functions, solving  the MQ problem is equivalent to invert the function $\PP: (x_1,\ldots,x_n)\longmapsto (p_1(x_1,\ldots,x_n),\ldots,p_m(x_1,\ldots,x_n))$, i.e. given $d=\PP(z)\in {\FF_q}^m$, it is unfeasible to recover $z\in {\FF_q}^n$ if it is not known the way in which the polynomials $p_i$ are generated.

In a multivariate public key cryptosystem (MPKC),
$\PP$ is obtained using a secret  set of $m$ quadratic polynomials with random coefficients $\{f_1(x_1,\ldots,x_n),\ldots,f_m(x_1,\ldots,x_n)\}$ and composing the quadratic map  $$\FFF:(x_1,\ldots,x_n)\longmapsto (f_1(x_1,\ldots,x_n),\ldots,f_m(x_1,\ldots,x_n))$$ with two affine maps $\SS$ and $\TT$. The quadratic map $\FFF$ is easy to invert \cite{DGS}, but the action of the affine maps makes difficult to invert $\PP$, because the polynomials $p_i$  induced by $\PP$ are approximately random \cite{B}.

A system induced from the previous construction is known as bipolar system \cite{DGS}, and it  determines  different MPKCs depending on the particular choice of $\SS,\FFF$ and $\TT$.

\subsection{One-Way functions}
A one-way function is any function which can be efficiently computed but whose whose pseudo-inverse is hard to find. More formally:
\begin{defn}
A function $f$ is said to be one-way if it can be computed in polynomial time on any input $x$ and if any polynomial-time probabilistic algorithm used to solve $f(x)=y$ knowing $f$ and $y$ succeds with negligible probability.
\end{defn}
Examples of one-way functions are (cryptographically secure) block ciphers and hash functions. These primitives can therefore be safely used in the design of post-quantum digital signatures, since the only known speed-up for a quantum computer is Grover's search algorithm \cite{grover}, which however is not capable of determining in polynomial time a pre-image of a one-way function with more than negligible probability.

\subsection{Zero-Knowledge proofs}
Zero-knowledge proofs (ZKP) are protocols in which a prover can convince a verifier that a statement is true, without disclosing any information apart that the statement is true. The three classic properties that a ZKP needs are
\begin{itemize}
\item[$\bullet$] completeness: honest verifiers will be convinced by honest provers.
\item[$\bullet$] soundness: no malicious prover can prove (with non-negligible probability) a false statement.
\item[$\bullet$] zero-knowledge: no verifier learns anything other than the fact that the statement is true.
\end{itemize}

\section{Signature Schemes}\label{sec: schemes}
\subsection{Rainbow}

Rainbow \cite{Rainbow} is a generalisation of the Unbalanced Oil and Vinegar (UOV) signature scheme \cite{KPG}, obtained by considering multiple UOV layers. 
The security of Rainbow is linked to the NP-hard problem of solving a multivariate polynomial system of quadratic equations over the field $\FF=\mathbb{F}_{2^s}$.
The fundamental parameters of Rainbow are $s$, three positive integers $v_1$, $o_1$ and $o_2$, and a hash function $H$ whose digest is $(o_1+o_2) \cdot 2^s$ bits long.

Define two constants $m= o_1+o_2$ and $n=v_1+o_1+o_2=m+v_1$ and let $V_1=\{1,\ldots,v_1 \}$, $V_2=\{1,\ldots,v_1+o_1\}$, $O_1=\{v_1+1,\ldots\ ,v_1+o_1\}$ and $O_2=\{v_1+o_1+1,\ldots\ , n\}$ be four sets of integers determined by the parameters, and let $\SS: \FF^{m}\longrightarrow \FF^{m}$ and $\TT:\FF^{n}\longrightarrow \FF^{n}$ be two invertible affine maps. 
For each $k\in O_1\cup O_2$ define the map $f_k:\FF^n\to \FF$ according to the formula
\begin{align*}
f_k(x_1,\ldots,x_n)&=\sum_{i,j \in V_l, i\leq j} \alpha_{k,i,j}x_i x_j+ \sum_{i\in V_l, j\in O_l} \beta_{k,i,j}x_i x_j+\\&+ \sum_{i\in V_l \cup O_l} \gamma_{k,i,j} x_i +\delta_k\;,
\end{align*}
where $l\in\{1,2\}$ is the unique index for which $k\in O_l$ and $\alpha_{k,i,j}$, $\beta_{k,i,j}$, $\gamma_{k,i,j}$, $\delta_k \in \FF$ are randomly generated parameters.
The $m$ functions in $n$ variables $f_{v_1+1},\ldots,f_n$ are used to define a quadratic map 
$\FFF: \FF^{n}\longrightarrow \FF^{m}$, such that $${\FFF(x_1, \ldots,x_n)=(f_{v_1+1}(x_1, \ldots ,x_n),\ldots,f_n(x_1, \ldots ,x_n))}.$$
Due to the structure of $\FFF$, given $d\in\FF^m$ it is possible to find in a reasonable amount time a value $\bar{z}\in\FF^n$ such that $\FFF(\bar{z})=d$, employing an algorithm that fixes the first variables and then applies Gaussian elimination. This property is used to efficiently compute a value $z$ such that $\PP(z)=\SS\circ\FFF\circ\TT(z)=d$. On the other hand, given $z$ and $\PP$ it is easy to compute $d=\PP(z)$, but from $d\in\FF^m$ it is unfeasible to obtain a value $z\in\FF^n$ for which $\PP(z)=d$ without knowing $\FFF$ if $\SS, \TT$, and $\FFF$ are random. 

Given the parameters $(s,v_1,o_1,o_2,H)$ described above, the protocol works as follows.

\subsubsection{Key generation}
\begin{enumerate}
 \item Randomly choose $\SS$, $\TT$ and $\FFF$ as defined above, choosing the maps' coefficients uniformly at random in $\FF$.
 \item The private key consists of $(\SS,\FFF,\TT)$.
 \item The public key is the composition  $\PP=\SS \circ \FFF \circ \TT$.
\end{enumerate}

\subsubsection{Signing}
Given a key-pair $((\SS,\FFF,\TT),\PP)$ and a message digest $d$, compute the signature performing the following steps:
\begin{enumerate}
\item Choose uniformly at random a bit string $r$ with the same length of $d$.
\item Compute $h=H(d||r)$ interpreted as a vector of $\FF^m$.
\item Compute  $x=\SS^{-1}(h)$.
\item Compute $y=\FFF^{-1}(x)$.
\item Compute $z=\TT^{-1}(y)$.
\item The signature is the pair $(z,r)$.
\end{enumerate}

\subsubsection{Verification}
To verify a signature $(z,r)$ on a message digest $d$ perform the following steps:
\begin{enumerate}
\item Compute $h=H(d||r)$ interpreted as a vector of $\FF^m$.
\item Compute $h'=\PP(z)$ and check if $h'=h$.
\end{enumerate}

\subsection{GeMSS}
GeMSS \cite{GeMSS} is a multivariate signature scheme, based on a  system of polynomial equations over the field $\FF_2$.
The fundamental parameters of GeMSS are the following: $m$ the number of equations, $\Delta$  and $v$ that determine the number of total variables, and a hash function $H$ whose digest is $k$ bits long.\\
Fix $n=m+\Delta$  and let $S\in \mathrm{GL}_{n+v}(\FF_2)$ and $T\in \mathrm{GL}_{n}(\FF_2)$ be two invertible matrices.
Define $F\in \FF_{2^n}[X,v_1, \ldots ,v_v]$, a polynomial of degree $D$, with the following structure:
\begin{align*} F(X,v_1, \ldots, v_v)&=\sum_{\substack{0\leq j<i<n \\  2^i+2^j\leq D}} A_{i,j}X^{2^i+2^j}+\sum_{\substack{0\leq i<n \\  2^i\leq D}} \beta_{i}(v_1,\ldots,v_v)X^{2^i}+\\ &+ \gamma(v_1, \ldots, v_v), \end{align*}
where $A_{i,j}\in \FF_{2^n}$, each $\beta_i : {\FF_2}^{v} \longrightarrow {\FF_2}^n$ is linear and $\gamma(v_1,\ldots , v_v) : {\FF_2}^{v} \longrightarrow {\FF_2}^{n}$ is quadratic.\\
Let $(\theta_1, \ldots, \theta_n)\in {\FF_{2^n}}^{n}$ be a basis of ${\FF_2}^{n}$ over $\FF_2$. Given $E= \sum_{k=1}^{n}e_k \cdot \theta_k \in \FF_{2^n}$, define the following function :
\begin{equation} \Phi:\FF_{2^n} \longrightarrow {\FF_2}^{n} \ \ \ E \longmapsto \Phi(E)=(e_1,\ldots,e_n).  \label{phi}
\end{equation}
Starting from $F$, it is possible to define $n$ multivariate polynomials $f_k \in \FF_2[x_1, \ldots, x_{n+v}]$, such that $F\big( \sum_{k=1}^{n} \theta_k x_k,v_1 \ldots,v_v \big)=\sum_{k=1}^{n} \theta_k f_k$.
The public key $P$ is derived from $f_1,\ldots,f_n$ and it consists of the first $m$ components of 
\begin{equation}(p_1, \ldots, p_n)=\left(f_1\left(\left(x_1,\ldots,x_{n+v}\right)\cdot S\right), \ldots,f_n\left(\left(x_1,\ldots,x_{n+v}\right)\cdot S\right)\right) \cdot T,
\label{pk}
\end{equation}
which is reduced modulo the field equations, that is \linebreak[4]${\mathrm{mod} \langle x_1^2-x_1, \ldots, x_n^2-x_n \rangle}$.  
Due to the structure of $F$, given $d\in {\FF_2}^{m}$ and $r\in {\FF_2}^{n-m}$ randomly chosen, it is possible, with a procedure that fixes the last $v$ variables and then applies Berlekamp's algorithm on the resulting univariate polynomial, to find a root of $F-\Phi^{-1}((d,r) \cdot T^{-1})$ in a reasonable  amount time $(O(nD))$. This property is used to efficiently compute a value $z\in {\FF_2}^{n+v}$ such that $P(z)=d$. On the other hand, given $z$ and $P$ it is easy to compute $d=P(z)$, but from $d\in {\FF_2}^{m}$ it is unfeasible to obtain a value $z\in {\FF_2}^{n+v}$ for which $P(z)=d$ without knowing $F$, if $S$,$T$ and $F$ are random.\\
Finally,  it is possible to iterate $t$ times a part of the signature to increase the  security level $\lambda$, indeed in this way it is possible to apply the hash function $H$ and at the same time to combine the actions of $S$  and $T$ on the variables more than once.

 Given the parameters $(m,\Delta,v,D,H,t)$ described above, the protocol works as follows.
\subsubsection{Key generation}
\begin{enumerate}
    \item Randomly choose $S,T$ and $F$ choosing the coefficients of $F$ uniformly at random in $\FF_{2^{n}}$ and the elements of $S$ and $T$ in $\FF_2$.
    \item The private key consists of $(S,T,F)$.
    \item Compute $p=(p_1,\ldots,p_n)$ as defined in ~\eqref{pk}.
    \item The public key is $P=(p_1,\ldots,p_m)$, the first $m$ components of $p$.
\end{enumerate}
\subsubsection{Signing}
Given a key-pair $((S,T,F),P)$ and a message digest $h$, compute the signature performing the following steps:
\begin{enumerate}
   \item Set $S_0=0\in {\FF_2}^{m}$.
   \item Repeat for $i=1$ to $t$ the following steps:
   \begin{enumerate}
       \item Get $D_i$ the first $m$ bits of $h$ and compute $D'_i=D_i \oplus S_{i-1}$.
       \item Randomly choose $(v_1,\ldots,v_v)\in {\FF_2}^{v}$ and $r\in {\FF_2}^{n-m}$.
       \item Compute $A_i=\phi^{-1}((D'_i,r) \cdot T^{-1})$ as described in ~\eqref{phi}.
       \item Compute a root $Z$ of $F-A_i$.
       \item Compute $(S_i,X_i)=(\phi(Z),v_1,\ldots,v_v)\cdot S^{-1}\in {\FF_2}^{m} \times {\FF_2}^{n+v-m}$.
       \item Compute $h=H(h)$.
   \end{enumerate}
   \item The signature is $z=(S_{t},X_{t}, \ldots, X_1)$.
\end{enumerate}
\subsubsection{Verification}
To verify a signature $z$ on a message digest $h$ perform the following steps:
\begin{enumerate}
\item Repeat for $i=1$ to $t$
\begin{enumerate}
    \item Get $D_i$ the first $m$ bits of $h$.
    \item Compute $h=H(h)$.
\end{enumerate}
\item Repeat for $i=t-1$ to $0$
 \begin{enumerate}
     \item Compute $S_i=P(S_{i+1},X_{i+1})\oplus D_{i+1}$.
 \end{enumerate}
\item Check if $S_0=0$.
\end{enumerate}

\subsection{CRYSTALS-DILITHIUM}
CRYSTALS-DILITHIUM \cite{Crystals} is a lattice-based signature built on the hardness of two problems: MLWE and SelfTargetMSIS problem~\cite{KLS}, a variation of the MSIS problem.
The first problem is defined over a polynomial ring $R_q= \mathbb{Z}_q[x]/(x^{256}+1)$, where $q$ is a prime such that $q\equiv 1 \mod 512$.
This condition on $q$ allows to use the NTT (Number Theoretic Transform, a generalization of the discrete Fourier transform over a finite field) representation.
Given $H$ a hash function and a vector $x$,
 SelfTargetMSIS consists in finding a vector $z'=(z,c)$  with small coefficients such that $H(x||f(z'))=c$ with $\mathtt{w}(c)=60$ (where $\mathtt{w}$ denotes the Hamming weight), where $f$ is a linear function.\\
 In order to define an ordering relation in $\mathbb{Z}_q$, we will consider the embedding $\eta: \mathbb{Z}_q \to \ZZ$, where $\eta(z) \equiv z \bmod{q}$ and $ -\frac{q-1}{2} \leq \eta(z) \leq  \frac{q-1}{2}$. For any $z_1$, $z_2 \in \ZZ_q$ we say that $z_1 \leq z_2$ if and only if $\eta(z_1) \leq \eta(z_2)$.

Let  $w=w_0+w_1x+\ldots+w_{255}x^{255}$ be a polynomial in $R_q$, the norm $\Vert{w}\Vert_\infty:=\displaystyle\max_{i}(|w_i|)$ is used to check some conditions related to the security and correctness, for this reason it is introduced a parameter $\beta\in\mathbb{Z}$ as a bound for the norm of some quantities.
Let $A \in M_{k,l}(R_q)$ be a matrix and set $\bar{w}=Ay$, where $y\in R_q^l$ is a vector such that $\Vert{y}\Vert_\infty\leq \gamma_1 $ (with $\gamma_1\in\mathbb{Z}$ another parameter),  we  distinguish between the high-order and low-order parts of $\bar{w}$ as follows: for each component $w'$ of $\bar{w}$
\begin{equation}\label{eqw}
    w'=w'_1 \cdot 2\gamma_2+w'_0,
\end{equation}
where $\Vert{w'_0}\Vert_\infty\leq\gamma_2$, where $\gamma_2\in \mathbb{Z}$ is another parameter.
We call $w'_1$ the high-order part, while $w'_0$ is the low-order part of $w'$.
We denote with $\mathrm{HB}(\bar{w})$ (HighBits) the vector comprising all $w'_1$s, thus is the high-order part of $\bar{w}$, and with $\mathrm{LB}(\bar{w})$ (LowBits) the low-order part of $\bar{w}$.

For storage efficiency, instead of generating and storing the entire matrix $A$, the protocol makes use of a secure PRNG and the NTT: using the NTT, it is possible to identify $a\in R_q$ and $\bar{a}\in \mathbb{Z}_q^{256}$, where $\bar{a}$ is the NTT representation of $a$, while if $A\in R_q^{k\times l}$ we denote with $\ntt(A)$ the matrix where each
coefficient of $A$ is identified with an element of ${\mathbb{Z}_q}^{256}$.
To obtain the matrix $A$, every  element $\bar{a}_{i,j}$ of $\ntt(A)$ is generated from a $256$ bit random seed $\rho$.

The parameters of CRYSTALS-DILITHIUM are $q$, $d$, $k$, $l$, $\eta$, $\gamma_1$, $\gamma_2$, $\Omega$, $H$, $G$, where $\gamma_1, \gamma_2, k, l$ and $q$ are defined as above, $d \in \ZZ_q$, $\eta$ and $\Omega$ are other bounds and $G, H$ are hash functions.

\noindent Given the parameters $(q, k, l, \eta, G, H, d, \gamma_1, \gamma_2, \beta, \Omega)$ described above, the protocol works as follows.
\subsubsection{Key generation}
\begin{enumerate}
     \item Choose uniformly at random two bit strings $\rho$ and $\theta$ of length 256.
     \item Choose uniformly at random $(s_1,s_2)\in {R_q}^l \times {R_q}^k$ with $|s_i|\leq \eta$.
     \item Compute $A\in R_q^{k\times l}$ from $\rho$ using NTT representation.
     \item Compute $t=As_1+s_2$.
     \item Compute $t_0=t \ \mathtt{mod} \ 2^d$ and $t_1=\frac{t-t_0}{2^d}$.
     \item The public key is $P=(\rho,t_1)$.
     \item The private key is $S=(\rho,\theta,G(\rho||t_1),s_1,s_2,t_0)$.
\end{enumerate}

\subsubsection{Signing}
Given a key-pair $(S,P)$ and a message $M$ compute the signature performing the following steps:
\begin{enumerate}
    \item Compute $A$ from $\rho$ as described above.
    \item Compute $\mu=G(G(\rho||t_1)||M)$ and $\rho'=G(\theta||\mu)$.
    \item Compute uniformly at random $y\in R_q^l$ with $\Vert{y}\Vert_\infty<\gamma_1$, starting from seed $\rho'$ using NTT representation.
    \item Compute $w=Ay$ and $w_1=\HB(w)$.
    \item Compute $c=H(\mu||w_1)$ and $z=y+cs_1$.
    \item Compute $r_1=\HB(w-cs_2)$ and $r_0=\LB(w-cs_2)$.
    \item Check if all the following conditions are satisfied else repeat from step $3$:
    \begin{enumerate}
        \item $\Vert{z}\Vert_\infty<\gamma_1-\beta$.
        \item $\Vert{r_0}\Vert_\infty<\gamma_2-\beta$.
        \item $r_1=w_1$.
    \end{enumerate}
    \item Compute $h=(h_1,\ldots,h_k)=r_1\oplus \HB(w-cs_2+ct_0)$.
    \item Compute $\Omega'= \mathtt{w}(h)$  and check if $\Omega' \leq \Omega$ else repeat from step 3.
    \item The signature is $(z,h,c)$.
\end{enumerate}

\subsubsection{Verification}
To verify a signature $(z,h,c)$ on a message $M$ perform the following steps:
\begin{enumerate}
    \item Compute $A$ and $\mu$ as described in the signing process.
    \item Compute
    $w'_1=\HB(w-cs_2)$ knowing
    $\HB(A z - c t_1\cdot 2^d)=\HB(w - c s_2  +c t_0)$ and $h$ that allows to remove the error generated by $ct_0$.
    \item Check if all the following conditions are satisfied:
    \begin{enumerate}
    \item$\Vert{z}\Vert_\infty<\gamma_1-\beta$.
    \item $c=H(\mu||w'_1)$.
    \item Compute $\Omega'=\mathtt{w(h)}$ and check if $\Omega' \leq \Omega$.
    \end{enumerate}
\end{enumerate}

Given the parameter $d\in \mathbb{Z}_q$ and computed $z=y+cs_1$ with $s_1  \in R_q^l$, it is possible to define $t_1,t_0\in\mathbb{Z}_{q}$ such that $t=t_1 \cdot 2^d +t_0$ ($t_1$ is the high order part of $t$) and compute $\HB(\bar{w}-cs_2+ct_0)$. Indeed:
$$Az-ct_1\cdot2^d=Ay+cAs_1-c(t-t_0)=Ay-cs_2+ct_0=\bar{w}-cs_2+ct_0.$$
Starting from $r_1=\HB(\bar{w}-cs_2+ct_0)$, it is easy to obtain $\HB(\bar{w}-cs_2)$ knowing $h=r_1\oplus \HB(\bar{w}-cs_2)$, indeed it is sufficient  to check which bits of $h$ have value $1$ to find the error bits in $r_1$ and  changing their value. The arithmetic modulus $\frac{q-1}{2\gamma_2}$ is required to modify $r_1$ depending on the sign of $\LB(\bar{w}-cs_2+ct_0)$. Besides, the parameter $\Omega$ is the maximum Hamming weight that $h$ can assume and thanks to the condition $\Vert{z}\Vert_\infty<\gamma_1-\beta$, it is possible to make the correction of error bits successfully, in a safe way. On the other hand, it is infeasible to recover $z$ without knowing $y$ (so $\bar{w}$ cannot be computed) and $s_1$.

\subsection{FALCON}
FALCON \cite{FALCON} is a particular lattice-based signature, which is based on solving the SIS problem over the NTRU lattices. Given $n=2^k$, $q\in \mathbb{N}^*$ and defined $R$ using  $\phi(x)=x^n+1$, the problem consists in determining  $f,g,G,F\in R$ such that $f$ is invertible modulus $q$ (this condition is equivalent to require that $\mathrm{NTT}(f)$ does not contain $0$ as a coefficient) and such that the following equation (NTRU equation) is satisfied:
\begin{equation}\label{NTRU}
fG-gF=q \mod \phi.
\end{equation}

If $h:=g \cdot f^{-1} \ \mathrm{mod} \ q$, it is possible to verify that the matrices $P=\begin{bmatrix} 
    1 & h \\
    0 & q 
\end{bmatrix}$ and $Q=\begin{bmatrix} 
    f & g \\
    F & G 
\end{bmatrix}$ generate the same lattice: $$\Lambda(P)=\{zP \ | \ z\in R_q \}=\{zQ \ | \ z\in R_q \}=\Lambda(Q),$$ but, if $f$ and $g$ are sufficiently small, then $h$ should seem random, so, given $h$, the hardness of this problem consists of finding $f$ and $g$.
Each coefficient of the polynomials $f=\displaystyle\sum_{i=0}^{n-1}f_i x^i$ and $g=\displaystyle\sum_{i=0}^{n-1}g_i x^i$ is generated from  a distribution close to a Gaussian of center $0$ and standard deviation $\sigma\in [\sigma_{\mathrm{min}},\sigma_{\mathrm{max}}]$ (where $\sigma,\sigma_{\mathrm{min}},\sigma_{\mathrm{max}}$ are parameters).\\
The following general property is fundamental to solve the NTRU equation~\ref{NTRU}, in particular if $f=\displaystyle\sum_{i=0}^{n-1}a_ix^i\in \mathbb{Q}[x]$, $f$ can be decomposed in a unique way as:
\begin{equation}\label{dec}
f(x)=f_0(x^2)+xf_1(x^2),
\end{equation} where $f_0=\displaystyle\sum_{i=0}^{n/2-1}a_{2i} x^i$ and $f_1=\displaystyle\sum_{i=0}^{n/2-1}a_{2i+1} x^i$.\\
Given $f$ and $g$, it is easy to obtain $(F,G)$ solution of \eqref{NTRU}, indeed there is a recursive procedure that uses the previous property and allows to solve a NTRU equation  in the ring $\mathbb{Z}=\mathbb{Z}[x]/(x+1)$ and then transforms this solution $(F,G)\in \mathbb{Z} \times \mathbb{Z}$ into two polynomials of $\mathbb{Z}[x]/(\phi)$.
Thanks to the FFT, it is possible to define the matrix $\bar{B}=\begin{bmatrix} 
    \mathrm{FFT}(g) & \mathrm{FFT}(-f) \\
    \mathrm{FFT}(G) & \mathrm{FFT}(-F) 
\end{bmatrix}$. Moreover we also need to consider the LDL decomposition of  $\mathcal{G}=\bar{B}\cdot\bar{B}^{T}=LDL^{T}$, where $L=\begin{bmatrix} 
    1 & 0 \\
   \bar{L} & 1
\end{bmatrix}$
and $D=\begin{bmatrix} 
    D_{11} & 0 \\
   0 & D_{22}
\end{bmatrix}$. \\
Starting from $\mathcal{G}\in M_{2,2}\left(\mathbb{Q}[x]/(\phi)\right)$, it is possible to construct the so-called FALCON tree $T$: the root of $T$ is $\bar{L}$ and its two child-nodes  $G_0, G_1\in M_{2,2}\left(\mathbb{Q}[x]/\left(x^{n/2}+1\right)\right)$ are obtained considering the decomposition of $D_{11}$ and $D_{22}$ as described in \eqref{dec}. Iterating this procedure on $G_0$ and $G_1$, it is possible to obtain the whole tree $T$, where each leaf $l\in \mathbb{Q}$ is normalized, i.e $l'=\frac{\sigma}{l}$.\\
 FALCON uses a particular  hash function $H$, which transforms a string modulus $q$ in a polynomial $c\in \mathbb{Z}_q[x]/(\phi)$. \\ In addition to the standard deviations described above, the parameters of FALCON are  $k$, $q$ and two other constants $\beta\in \mathbb{Q_+} $ and $b_l\in\mathbb{N}^{*}$ that  will be described later.\\
 Given a solution $t$ of $\bar{B} t=c$, there exists a recursive procedure (Fast Fourier sampling), which applies a randomized rounding on the coefficients of $t\in \mathbb{Q}[x]/(\phi)$ to obtain a polynomial $z\in  R$, using the information stored in $T$.\\
Let $a,b\in \mathbb{Q}[x]/(\phi)$, it is possible define the following inner product and its associated norm: $$<a,b>=\frac{1}{n}\displaystyle\sum_{i\in \mathbb{C} \;:\; \phi(i)=0}a(i)\cdot \displaystyle\overline{b(i)}. $$
Let $\beta\in \mathbb{Q_+}$, then it is possible to compute $s=(t-z)\bar{B}$ with $|| s||^2\leq \left\lfloor \beta^2 \right\rfloor$  and using the inverse of FFT it is easy to compute $s_1,s_2\in R$, which satisfy: 
\begin{equation}\label{firma}
 s_1+s_2h=c \ \mathrm{mod} \ q.   
\end{equation}
On the other hand, given $(s_1,s_2)$ it is unfeasible to recover $s$ without knowing $\bar{B}$ and $T$.\\
Finally, FALCON  uses a compression algorithm, which transforms $s_2$ in a byte string  $(8 \cdot b_l  -328)$ long.

 Given the parameters $(k,q,\sigma_{\mathrm{min}},\sigma_{\mathrm{max}},\sigma,\beta,b_l)$ described above, the protocol works as follows.
\subsubsection{Key generation}
\begin{enumerate}
    \item Compute $f=\displaystyle\sum_{i=0}^{n-1}f_i x_i$ and $g=\displaystyle\sum_{i=0}^{n-1}g_i x_i$, generating $f_i$ and $g_i$ from Gaussian distribution $D_{0,\sigma}$.
    \item Check that $f$ is invertible modulus $q$, else restart from step 1.
    \item Find $(F,G)$ solution of the NTRU equation \eqref{NTRU}.
    \item Compute $\bar{B}$ as described above.
    \item Compute $G=\bar{B}\cdot\bar{B}^{T}$, obtain the FALCON tree $T$ using LDL decomposition and normalize its leaves.
    \item Compute $h=gf^{-1} \ \mathrm{mod} \ q$.
    \item The private key is $(\bar{B},T)$.
    \item The public key is $h$.
\end{enumerate}

\subsubsection{Signing}
Given a key-pair $(h,(\bar{B},T))$ and a message $m$, compute the signature performing the following steps:
\begin{enumerate}
    \item Choose uniformly at random a bit string $r$ 320 long.
    \item Compute $c=H(r||m,q,n)$ and solve $\bar{B}t=c$.
    \item Compute $z$ randomized rounding of $t$ as described above.
    \item Compute $s=(t-z)\bar{B}$.
    \item Check that $|| s||^2\leq\lfloor{\beta^2}\rfloor$ else repeat from step 3.
    \item Compute $(s_1,s_2)$ satisfying \eqref{firma}.
    \item By compressing $s_2$, compute a string $s'$ of $(8\cdot b_l-328)$ bytes.
    \item The signature is $(r,s')$.
\end{enumerate}
\subsubsection{Verification}
Given a public key $h$, to verify a signature $(r,s')$ on a message digest $c$ perform the following steps: 
\begin{enumerate}
    \item By decompressing $s'$, compute $s_2$.
    \item Compute $s_1=c-s_2h \ \mathrm{mod} \ q$.
    \item Check if $|| (s_1,s_2)||^2\leq\lfloor{\beta^2}\rfloor$.
\end{enumerate}

\subsection{\texorpdfstring{SPHINCS$^+$}{SPHINCS+}}\label{sphincs}
SPHINCS$^+$ \cite{Sphincs+} is based on hash functions and it is nothing more than an opportune union of three signature schemes: WOTS$^+$ \cite{WOTS}, XMSS \cite{XMSS} and FORS \cite{FORS}.\\
SPHINCS$^+$ works with two main tree structures: a Hypertree and a FORS tree.
The Hypertree consists  of $d$ Merkle trees of height $h'$. On each of these trees is applied an XMSS signature scheme. XMSS, in turn, consists of a one-time signature WOTS$^{+}$ applied on the root of the previous layer plus the authentication path of the randomly chosen leaf.\\
On the other hand, a FORS tree is made up of $k$ parallel trees of height $a$ and, contrary to Hypertrees, this kind of trees is used only on signature generation and verification, but not for key generation.\\
SPHINCS$^{+}$ uses the FORS scheme to generate a hash value that relates the message to $k$ FORS roots.
After that, a Hypertree signature is applied to the hash returned by the FORS signature to generate a SPHINCS$^{+}$ signature.

The security of this scheme derives from the security of the hash function involved. In particular SPHINCS$^{+}$ uses the so called \emph{tweakable hash functions}, which allow us to approach the details of how exactly the nodes are computed.\\
The choice of the hash function strongly influences the security of the signature, in fact the length $n$ of every hash value of this protocol is fundamental to determinate the security level, the authors have chosen SHAKE256 as the hash function family.\\
The parameters $k$ and $a$ determine the performance and security of FORS, so  it is necessary to balance the value of these two parameters to avoid getting too large or too slow signatures.
Instead, the height of the Hypertree $h'd$ determines the number of XMSS instances, so this value has a direct impact on security: a taller Hypertree gives more security. Remark that the number of layers $d$ is a pure performance trade-off parameter and does not influence security. \\
Finally, the Winternitz parameter $w$ is a trade-off parameter (greater $w$ means shorter signatures but slower signing), which determines the number and length of the hash chains per WOTS$^{+}$ instance.\\
The privacy of SPHINCS$^{+}$ is guaranteed by the pseudorandom generation of WOTS$^{+}$ and FORS secret keys (this operation randomizes the choice of FORS and WOTS$^{+}$ leaves used to sign).

Given the parameters $(n,h',d,k,a,w)$ described above, the protocol works as follows.

\subsubsection{Key generation}
The description of key generation assumes the existence of a function secRand which on input $n$ returns $n$ bytes of cryptographically strong randomness.
\begin{enumerate}
\item Compute SK.seed=secRand($n$), which is used to generate all the WOTS$^+$ and FORS private key elements.
\item Compute SK.prf=secRand($n$), which is used to generate a randomization value for the randomized message hash.
\item Compute PK.seed=secRand($n$), which is the public seed.
\item Compute PK.root, which is the hypertree root, i.e. the XMSS root of the tree on the top level.
\item The private key is: SK=(SK.seed, SK.prf, PK.seed, PK.root).
\item The public key is: PK=(PK.seed, PK.root).
\end{enumerate}
\subsubsection{Signing}
Given the private key SK and a message M, compute the signature performing the following steps:
\begin{enumerate}
\item Compute R, an $n$-bytes string pseudorandomly generated starting from SK.prf and M.
\item Compute the digest of M.
\item Compute SIGFORS, which is a FORS signature applied to the first $ka$ bits of the digest.
\item Starting from SIGFORS, derive PKFORS i.e. the public key associated to the FORS signature.
\item Compute HTSIG, which is an hypertree signature applied to PKFORS.
\item The SPHINCS$^{+}$ signature is: SIG=(R, SIGFORS, HTSIG).
\end{enumerate}
\subsubsection{Verification}
To verify a signature SIG on a message M perform the following steps:
\begin{enumerate}
\item Get R, which corresponds to the first $n$ bytes of SIG.
\item Get SIGFORS, the following $k(a + 1) \cdot n$ bytes of SIG.
\item Compute the digest of  M.
\item Starting from SIGFORS and the first $ka$ bits of the digest, derive PKFORS.
\item Starting from PKFORS, check the hypertree verification.
\end{enumerate}

\subsection{Picnic}\label{picnic}
Picnic \cite{Picnic} is a signature scheme whose security is based on the \emph{one-wayness} of a block cipher and the pseudo-random properties of an extensible hash function.

In particular the construction relies upon the fact that a digital signature is essentially a \emph{non-interactive zero knowledge proof of knowledge} of the preimage of a one-way function output, where the \emph{challenge} inside the proof is tied to the message that is being signed.
In other words, the signer creates a transcript that demonstrates the knowledge of the private key whose image through the one-way function is the public key, without revealing any information about the private key itself. Moreover this transcript is indissolubly bound to the message.

Starting from this general idea, Picnic instantiates a signature scheme using classical general-purpose primitives: a block cipher, a secure multi-party computation protocol (MPC), and an extensible cryptographic hash function (also known as extensible output function or XOF).
The zero-knowledge proof (ZKP) is derived from the hash and the MPC protocol, exploiting the security properties of the latter.
The prover computes the one-way function using its multi-party decomposition, controlling every party.
The security of the MPC protocol allows the disclosure of the complete view of some (in this case all but one) parties without revealing anything about the secret input, so a ZKP may be constructed committing to every view and randomly selecting which ones to reveal (the challenge).
The commitment (built from the hash) \emph{binds} the prover to the views (i.e. they cannot be changed after the commitment) without revealing them yet (the commitment is \emph{hiding}).
\emph{Soundness} can be achieved repeating this process for a few iterations, so that the verifier can be convinced that the prover could not have successfully produced the views without actually knowing the MPC input, except with negligible probability.

The protocol just described is interactive, but there are fairly simple techniques that allow to transform it into a \emph{non-interactive} one, i.e. a transcript produced by the prover that by itself can convince a verifier.
These techniques use a deterministic pseudo-random generator (the hash) to derive the challenges from the public values, i.e. the public key, the commitments and the message.
Assuming the (quantum) random oracle model~\cite{ROM,QROM} (i.e. the hash is modeled through an oracle that outputs random values on new inputs, but does not change answer when a query is repeated), we maintain soundness even without interaction, and the message is tightly fastened to the transcript, so that it is infeasible to adapt this signature for another message without knowing the private key.

The MPC protocols are much more sensitive to the number of AND operations on two secret bits than to XOR operations, since the masking of AND gates requires extra information to keep consistency.
This, in turn, causes the MPC views (and thus the signature) to grow in size, therefore Picnic selected as block cipher LowMC~\cite{LowMC}, an algorithm designed to minimize such operations for a given security level.
LowMC employs a classic \emph{substitution-permutation} structure with $n$-bit blocks (where $n$ essentially defines the security level of the whole signature) and $r$ rounds in which $s$ parallel 3-bit S-boxes are applied (note that they do not necessarily cover the entire block), followed by a linear permutation (defined by a different matrix for each round), and a round-key addition (the round-keys are derived multiplying the key by $r+1$ different matrices: one for the initial key-whitening and again one per round).

Picnic's NIST submission defines various parameters sets that, besides optimizing LowMC parameters for the three security levels, employ different MPC protocols and techniques to obtain a non-interactive zero-knowledge proof (NIZKP).
More specifically, \texttt{picnic-LX-FS} (where $X\in\{1,3,5\}$ is the security level defined by NIST) uses the proof sistem ZKB++ (an optimized version of ZKBoo~\cite{ZKBoo}, a ZKP for boolean circuits based on an MPC called \emph{``circuit decomposition''}) that simulates $T$ parallel MPC executions between 3 parties, and uses the Fiat-Shamir transform~\cite{FS} to obtain a NIZKP.
The \texttt{picnic-LX-full} variant changes the LowMC parameters: uses a full S-box layer that allows to reduce the number of rounds.
The parameter sets \texttt{picnic-LX-UR} use again ZKB++ but with the Unruh transform~\cite{UR12,QROM,UR16}, which expands the signature size but is provable secure in the stronger quantum random oracle model (unlike the FS transform in general).
Finally the sets \texttt{picnic3-LX} bring along various optimizations: like \texttt{picnic-LX-full} they use a full S-box layer and the Fiat-Shamir transform, but they use a different ZKP and employ various optimizations to reduce signature size.
The ZKP used in \texttt{picnic3-LX} is the KKW protocol~\cite{KKW}, which simulates $T$ parallel MPC executions between $N$ parties ($N=16$ in the chosen parameters sets).
Each execution is divided into an offline preprocessing phase and an online phase where the shares are broadcast and the output reconstructed.
In KKW the challenger chooses $u$ executions for which the online phase will be revealed for all but one party, whereas for the other executions only the preprocessing phases will be revealed (for all parties).

Note that the MPC executions assume that each party consumes some random bits read from an input tape.
These tapes are deterministically generated from seeds through the XOF, and in turn those seeds are generated from a master seed, which is generated alongside a salt (used as extra input in every other derivation to prevent multi-target attacks such as in~\cite{MTattack}) from the secret key, the message, the public key, and the length parameter $S$ ( and optionally an extra random input to randomize signatures), always through the XOF.
The \texttt{picnic3-LX} parameters sets employ a tree structure to derive the seeds in order to reduce the amount of information needed to be included in the signature to reveal the MPC executions.
Moreover they use Merkle trees to compute the commitments, so the signatures can be compressed further.

All parameter sets use as XOF an instance of the SHAKE family~\cite{SHA3} (specifically SHAKE128 for security level L1 and SHAKE256 for L3 and L5) employing domain separation techniques to differentiate the uses as different random oracles.

Given the parameters $(S, n, s, r, T, u)$ described above, the protocol works as follows.
\subsubsection{Key generation}
\begin{enumerate}
    \item Choose a random $n$-bit string $p$, and a random $n$-bit string $k$.
    \item Using LowMC with parameters $(n, s, r)$, compute the encryption of $p$ with $k$, denoted $C = E(k, p)$.
    \item The private key is $k$.
    \item The public key is $(C, p)$.
\end{enumerate}
\subsubsection{Signing}
Given a key-pair $((C, p), k)$ and a message $M$, compute the signature performing the following steps:
\begin{enumerate}
    \item Derive the master seed and the salt from $k, M, (C, p), S$ (and possibly a random input of size $2S$), then derive the individual seeds.
    
    \item Simulate $T$ executions of the MPC protocols, producing for each party their view and output, starting from their seed.
    
    \item Compute the commitments to every seed and corresponding view.
    
    \item Compute the NIZKP challenge $e$ from the MPC outputs, the commitments, the salt, the public key and the message.
    
    \item Compute the NIZKP response by selecting for each MPC execution the appropriate outputs and decommitments to reveal, according to $e$.
    
    \item Assemble the signature $\sigma$ by including: $e$, the salt, the NIZKP response, and the commitments not derivable from the response.
\end{enumerate}

\subsubsection{Verification}
Given a public key $(C, p)$, to verify a signature $\sigma$ on a message $M$ perform the following steps:
\begin{enumerate}
    \item Deserialize $\sigma$ extracting the NIZKP challenge $e$, the salt, the NIZKP response, and the commitments.
    
    \item Parse the NIZKP response to obtain, for each of the $T$ MPC executions, the outputs and the decommitments prescribed by $e$.
    
    \item Use the seeds included in the decommitments to derive (with the salt) the tapes of the revealed parties, then use these values and the rest of the response to simulate the MPC executions that compute the LowMC encryption of $p$ with output $C$, computing the views for which the commitments are not included in the signature.
    
    \item Complete the commitments deriving the missing values from the results of the previous step, then derive the challenge $e'$ as in signing.
    
    \item The signature is valid if every parsing/deserialization succeeds, the MPC computations are correct, and $e' = e$.
\end{enumerate}

\section{Comparison}\label{sec: conclusions}
\begin{figure}[hbt!]
    \centering
    \includegraphics[width=\linewidth]{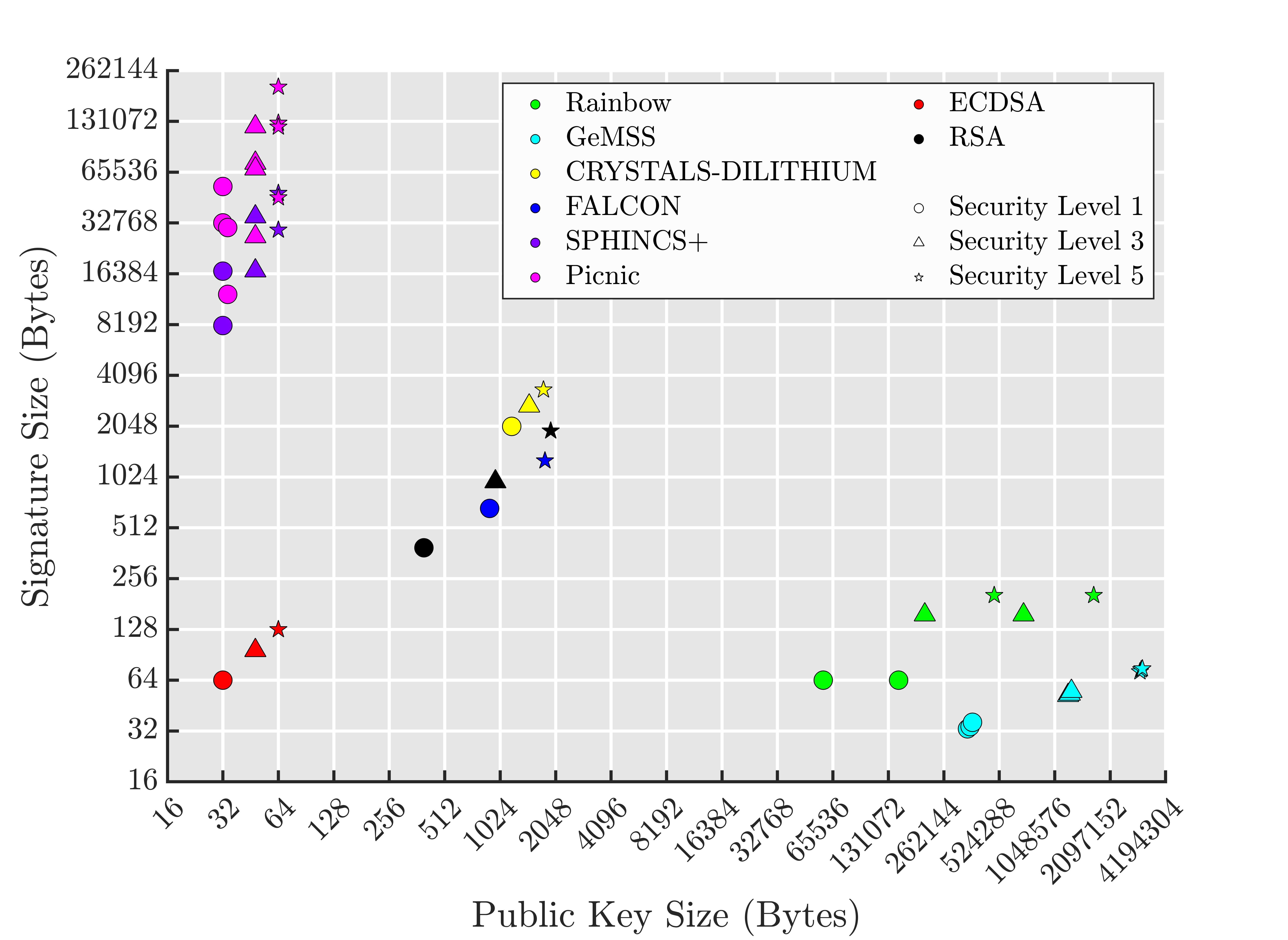}
    \caption{Signature-Public Key Size Comparison}
    \label{fig:sizes}
\end{figure}

In Figure \ref{fig:sizes} we summarize the dimensions in bytes of the public keys and the corresponding dimensions of the signatures of all the schemes presented in this survey, as well as those of the two classical schemes ECDSA and RSA. It is interesting to notice that the multivariate schemes have small signatures, but the size of their public keys is the largest among all the schemes. On the other hand, SPHINCS$^+$ and Picnic have small public keys, but large signatures, while the algorithms based on lattices have intermediate values in terms of both public keys and signatures. Finally it is worth to point out that, among all the schemes depicted, the best compromise in terms of dimension is still obtained by the non-quantum scheme of ECDSA.


\begin{thebibliography}{1}

\bibitem{LowMC} Albrecht, M.R., Rechberger, C., Schneider, T., Tiessen, T. and Zohner, M., Ciphers for MPC and FHE, In Annual International Conference on the Theory and Applications of Cryptographic Techniques (pp. 430-454), Springer, Berlin, Heidelberg, (2015).

\bibitem{Ajtai} Ajtai, M., Generating hard instances of lattice problems, In Proceedings of the twenty-eighth annual ACM symposium on Theory of computing (pp. 99-108),  (1996).

\bibitem{Sphincs+} Aumasson, J.P., Bernstein, D.J., Beullens, W., Dobraunig, C., Eichlseder, M., Fluhrer, S., Gazdag, S.L., H{\"u}lsing, A., Kampanakis, P., K{\"o}lbl, S., Lange, T., Lauridsen, M.M., Mendel, F., Niederhagen, R., Rechberger, C., Rijneveld, J., Schwabe, P., Westerbaan, B., {SPHINCS$^+$}. Submission to the {NIST} post-quantum project, v.3, \url{https://sphincs.org/data/sphincs+-round3-specification.pdf}, (2020). 

\bibitem{Crystals} Bai, S., Ducas, L., Kiltz, E., Lepoint, T., Lyubashevsky, V., Schwabe, P., Seiler, G., Stehlé, D., {CRYSTALS}-Dilithium Algorithm Specifications and Supporting Documentation. Round-3 submission to the {NIST PQC} project, \url{https://pq-crystals.org/dilithium/data/dilithium-specification-round3-20210208.pdf} (2021).

\bibitem{B} Bardet, M., \'{E}tude des syst\`{e}mes alg\'{e}briques surd\'{e}termin\'{e}s. Applications aux codes correcteurs et \`{a} la cryptographie These de doctorat de l'Universite Paris 6, Paris(2004)

\bibitem{ROM} Bellare M, Rogaway P., Random oracles are practical: A paradigm for designing efficient protocols. Proceedings of the 1st ACM Conference on Computer and Communications Security (pp. 62-73), (1993).

\bibitem{FORS} Bernstein, D. J., Hopwood, D., Hülsing, A., Lange, T., Niederhagen, R., Papachristodoulou, L., Schneider, M., Schwabe, P., Wilcox-O'Hearn, Z., SPHINCS: practical stateless hash-based signatures, In Annual international conference on the theory and applications of cryptographic techniques (pp. 368-397), Springer, Berlin, Heidelberg,  (2015). 

\bibitem{BGV} Brakerski, Z., Gentry, C., Vaikuntanathan, V., Fully homomorphic encryption without bootstrapping, ACM Transactions on Computation Theory (TOCT), 6(3), 1-36, (2014).

\bibitem{XMSS} Buchmann, J., Dahmen, E., Hülsing, A., {XMSS}-a practical forward secure signature scheme based on minimal security assumptions, In International Workshop on Post-Quantum Cryptography (pp. 117-129), Springer, Berlin, Heidelberg, (2011).

\bibitem{GeMSS} Casanova, A., Faugere, J.C., Macario-Rat, G., Patarin, J., Perret, L., Ryckeghem, J., GeMSS: a great multivariate short signature, Submission to the NIST's post-quantum cryptography standardization process, \url{https://www-polsys.lip6.fr/Links/NIST/GeMSS_specification_round2.pdf}, (2017).

\bibitem{Picnic} Chase, M., Derler, D., Goldfeder, S., Katz, J., Kolesnikov, V., Orlandi, C., Remacher, S., Rechberger, C., Slamanig, D., Wang, X., Zaverucha, G, The picnic signature scheme, Submission to NIST Post-Quantum Cryptography project, \url{https://github.com/microsoft/Picnic/blob/master/spec/design-v2.2.pdf}, (2020).

\bibitem{DGS} Ding, J., Gower, J., Schimdt, D., Multivariate Public Key Cryptosystems, Cincinnati: Springer(2006)

\bibitem{Rainbow} Ding, J., Chen, M.S., Kannwischer, M., Patarin, J., Petzoldt, A., Schmidt, D., Yang, B.Y., Rainbow - Algorithm Specification and Documentation, Submission to the NIST's post-quantum cryptography standardization process, \url{https://drive.google.com/file/d/1tcGC38SSkF_csxpzJpkM3qzfsWJq1ywL/view?usp=sharing}, (2020).

\bibitem{MTattack} Dinur, I., Kales, D., Promitzer, A., Ramacher, S. and Rechberger, C., Linear equivalence of block ciphers with partial non-linear layers: application to LowMC, In Annual International Conference on the Theory and Applications of Cryptographic Techniques (pp. 343-372), Springer, Cham, (2019).

\bibitem{FS} Fiat, A. and Shamir, A., How to prove yourself: Practical solutions to identification and signature problems, In Conference on the theory and application of cryptographic techniques (pp. 186-194), Springer, Berlin, Heidelberg, (1986).

\bibitem{FALCON} Fouque, P. A., Hoffstein, J., Kirchner, P., Lyubashevsky, V., Pornin, T., Prest, T., Ricosset, T., Seiler, G., Whyte, W., Zhang, Z. FALCON: Fast-Fourier lattice-based compact signatures over NTRU. Submission to the NIST's post-quantum cryptography standardization process, \url{https://falcon-sign.info/falcon.pdf}, (2020).

\bibitem{ZKBoo} Giacomelli, I., Madsen, J. and Orlandi, C., Zkboo: Faster zero-knowledge for boolean circuits, In 25th {usenix} security symposium ({usenix} security 16) (pp. 1069-1083), (2016).

\bibitem{grover} Grover, L. K., A fast quantum mechanical algorithm for database search, Proceedings, in Proceedings of the 28th Annual ACM Symposium on the Theory of Computing (p. 212), (1996).


\bibitem{Silverman} Hoffstein, J., Pipher, J., Silverman, J. H, An introduction to mathematical 
cryptography (Vol. 1), New York: Springer (2008).

\bibitem{WOTS} Hülsing, A., {W-OTS$^+$}-shorter signatures for hash-based signature schemes, in International Conference on Cryptology in Africa (pp. 173-188), Springer, Berlin, Heidelberg, (2013).

\bibitem{KKW} Katz, J., Kolesnikov, V. and Wang, X., Improved non-interactive zero knowledge with applications to post-quantum signatures, In Proceedings of the 2018 ACM SIGSAC Conference on Computer and Communications Security (pp. 525-537), (2018).

\bibitem{NIST1} Kerry, C., Gallagher, {FIPS PUB 186-4} Federal information processing standards publication digital signature standard (DSS),  National Institute of Standard and Technology, (2013).

\bibitem{KLS} Kiltz, E., Lyubashevsky, V., Schaffner, C., A concrete treatment of Fiat-Shamir signatures in the quantum random-oracle model, In Annual International Conference on the Theory and Applications of Cryptographic Techniques (pp. 552-586), Springer, Cham, (2018).

\bibitem{KPG} Kipnis, A., Patarin, J., Goubin, L.,  Unbalanced oil and vinegar signature schemes, In International Conference on the Theory and Applications of Cryptographic Techniques (pp. 206-222), Springer, Berlin, Heidelberg, (1999). 

\bibitem{KR} Kreuzer, M., Robbiano, L., Computational commutative algebra 1. Springer-Verlag, Berlin
(2000).

\bibitem{LS} Langlois, A., Stehlé, D.,  Worst-case to average-case reductions for module lattices, Designs, Codes and Cryptography, 75(3), 565-599, (2015).

\bibitem{LM} Lyubashevsky, V.,  Micciancio, D., Generalized compact knapsacks are collision resistant, In International Colloquium on Automata, Languages, and Programming (pp. 144-155), Springer, Berlin, Heidelberg,  (2006, July).

\bibitem{LPR} Lyubashevsky, V., Peikert, C., Regev, O., On ideal lattices and learning with errors over rings, In Annual International Conference on the Theory and Applications of Cryptographic Techniques (pp. 1-23), Springer, Berlin, Heidelberg, (2010).

\bibitem{Menezes} Menezes, A., Van Oorschot, P., Vanstone, S., Handbook of applied cryptography, CRC press, (2018). 

\bibitem{MR1} Micciancio, D.,  Regev, O., Worst-case to average-case reductions based on Gaussian measures, SIAM Journal on Computing, 37(1), 267-302, (2007).

\bibitem{MR2} Micciancio, D.,  Regev, O., Lattice-based cryptography, In Post-quantum cryptography (pp. 147-191), Springer, Berlin, Heidelberg,  (2009).

\bibitem{NIST-stand} NIST, Post-Quantum Cryptography Standardization, \url{https://csrc.nist.gov/Projects/post-quantum-cryptography/post-quantum-cryptography-standardization}, Accessed: 2021-04-01

\bibitem{SHA3} NIST, SHA-3 Standard: Permutation-Based Hash and Extendable-Output Functions, National Institute of Standards and Technology (NIST), FIPS PUB 202, U.S. Department of Commerce, (2015).

\bibitem{PG} Patarin J., Goubin L., Trapdoor One-Way Permutations and Multivariate Polynomials, Proceedings of the First International Conference on
Information and Communication Security, LNCS 1334, 1997, pp. 356-368.

\bibitem{Peikert} Peikert, C., Public-key cryptosystems from the worst-case shortest vector problem, In Proceedings of the forty-first annual ACM symposium on Theory of computing (pp. 333-342),  (2009). 

\bibitem{Peikert2} Peikert, C., A decade of lattice cryptography, Foundations and Trends in Theoretical Computer Science, 10(4), 283-424, (2016).

\bibitem{PR} Peikert, C., Rosen, A., Efficient collision-resistant hashing from worst-case assumptions on cyclic lattices, In Theory of Cryptography Conference (pp. 145-166), Springer, Berlin, Heidelberg, (2006).

\bibitem{Regev1} Regev, O., On lattices, learning with errors, random linear codes, and cryptography, Journal of the ACM (JACM), 56(6), 1-40,  (2009).

\bibitem{Regev2} Regev, O., The learning with errors problem, Invited survey in CCC, 7(30), 11, (2010).

\bibitem{Regev3} Regev, O.,  On lattices, learning with errors, random linear codes, and cryptography, Journal of the ACM (JACM), 56(6), 1-40, (2009).

\bibitem{Stinson} Stinson, D. R., Paterson, M., Cryptography: theory and practice, CRC press, (2018).

\bibitem{UR12} Unruh, D., Quantum proofs of knowledge, In Annual international conference on the theory and applications of cryptographic techniques (pp. 135-152), Springer, Berlin, Heidelberg, (2012).

\bibitem{QROM} Unruh, D., Non-interactive zero-knowledge proofs in the quantum random oracle model, In Annual International Conference on the Theory and Applications of Cryptographic Techniques (pp. 755-784), Springer, Berlin, Heidelberg (2015).

\bibitem{UR16} Unruh, D., Computationally binding quantum commitments, In Annual International Conference on the Theory and Applications of Cryptographic Techniques (pp. 497-527), Springer, Berlin, Heidelberg, (2016).


\end{thebibliography}
\end{document}